\documentclass[onecolumn,preprintnumbers,amsmath,amssymb,superscriptaddress]{revtex4-2}
\usepackage{epsf}
\usepackage{graphicx}
\usepackage{sidecap}
\usepackage{array}
\usepackage{amsmath}
\usepackage[utf8]{inputenc}
\usepackage{amssymb}
\usepackage{dcolumn}
\usepackage{epstopdf}
\usepackage{bm}

\def\nn{\nonumber}

\def\beq{\begin{eqnarray}}
\def\eeq{\end{eqnarray}}

\renewcommand{\v}[1]{\ensuremath{\mathbf{#1}}} 
 
\let\baraccent=\= 
\renewcommand{\=}[1]{\stackrel{#1}{=}} 

\begin{document}


\title{Ultrafast relaxation of acoustic and optical phonons in a topological nodal-line semimetal ZrSiS}

\author{Yangyang Liu}\affiliation{Department of Physics, University of Central Florida, Orlando, Florida 32816, USA}
\author{Gyanendra Dhakal} \affiliation{Department of Physics, University of Central Florida, Orlando, Florida 32816, USA}
\author{Anup Pradhan Sakhya}\affiliation{Department of Physics, University of Central Florida, Orlando, Florida 32816, USA}
\author{John E. Beetar}\affiliation{Department of Physics, University of Central Florida, Orlando, Florida 32816, USA}
\author{Firoza Kabir}\affiliation{Department of Physics, University of Central Florida, Orlando, Florida 32816, USA}
\author{Sabin Regmi}\affiliation{Department of Physics, University of Central Florida, Orlando, Florida 32816, USA}
\author{Dariusz Kaczorowski}
\affiliation{$^{2}$Institute of Low Temperature and Structure Research, Polish Academy of Sciences, ul. Okólna 2, 50-422 Wrocław, Poland}
 \author{Michael Chini}\affiliation{Department of Physics, University of Central Florida, Orlando, Florida 32816, USA}, \affiliation{$^{3}$CREOL, the College of Optics and Photonics, University of Central Florida, Orlando, Florida 32816, USA}
  \author{Benjamin M. Fregoso$^\dagger$} \affiliation{$^{4}$Department of Physics, Kent State University, Kent, Ohio 44242, USA}
\author{Madhab Neupane$^\dagger$}\affiliation{Department of Physics, University of Central Florida, Orlando, Florida 32816, USA}


\date{\today}

\begin{abstract}
Despite being the most studied nodal-line semimetal, a clear understanding of the transient state
relaxation dynamics and the underlying mechanism in ZrSiS is lacking. Using time- and angle-resolved
photoemission spectroscopy, we study the ultrafast relaxation dynamics in ZrSiS and
reveal a unique relaxation in the bulk nodal-line state which is well-captured by a simple model based
on optical and acoustic phonon cooling. We find linear decay processes for both optical and
acoustic phonon relaxations with acoustic cooling suppressed at high temperatures. Our results
reveal different decay mechanisms for the bulk and surface states and pave a way to understand the
mechanism of conduction in this material.
 \end{abstract}

\maketitle

In  topological semimetals,  valence and conduction bands cross each other in the momentum space with the band crossing protected by crystal and/or time-reversal symmetries resulting in intriguing electronic and transport properties \cite{Neupanereview, Hasan, Zhangreview, Bansil, Dai}. Topological Dirac and Weyl semimetals are characterized by band crossing between bulk conduction and valence bands at discrete k-points in the Brillouin zone (BZ) \cite{Hasan_weyl, Armitage}, whereas, topological nodal line semimetals (TNSMs) are characterized by the band-crossings along a line or a ring \cite{Burkov, Neupane,Schoop,  Chiu, Gatti, Klemenz, Feng}.  
TNSMs have been theoretically \cite{Burkov, Yu, Xu, Yang, Kim, Chan} and experimentally explored in diverse material systems \cite{Bian-NC, Neupane,Schoop, Ilya_science, SrAs3, npj_drumhead, hfp2}. 
The extension of band crossing to a line or a ring in TNSMs leads to an increase in the density of states of carriers in these bands, thus providing greater carrier density with which the photons can interact \cite{kirby}. The increased density of Dirac fermions in TNSMs are favorable for the presence of electron correlation effects \cite{Balents}. 

ZrXY family of materials where X = Si, Sn, Ge and Y = S, Se, Te are among the most studied TNSMs due to their high crystal quality and flexibility of the chemical composition \cite{Neupane, Schoop, Rudenko, topp, hu, zrsix, chen,  lou,  hu2,   zhang, zrgete, zrgete2,YLChen-PRB, fu, Wang}. Among the ZrXY family of materials, ZrSiS is an especially interesting material, since it possesses linearly dispersing bands extending up to 2 eV, which are free from interference from trivial bands making this material an excellent prospect for studying Dirac physics \cite{Rudenko}. Even though several experimental studies have been reported on ZrSiS such as angle-resolved photoemission spectroscopy (ARPES) \cite{Neupane, zrsix, Schoop, YLChen-PRB, fu}, high-field magnetotransport measurements \cite{Singha, Sankar},  scanning tunneling microscopy \cite{STM}, frequency-independent optical conductivity \cite{Schilling}, high carrier mobility \cite{Matusiak, Sankar}, the experimental and theoretical description of the effect of phonon scattering in electron cooling processes have not been studied so far. Using time-resolved ARPES (tr-ARPES), it has been observed that the electronic correlations in ZrSiSe can be reduced by optical excitation of high-energy electron-hole pairs, which screens the Coulomb interaction thus leading to renormalization of the Dirac quasiparticles \cite{Gatti}. A recent study on ultrafast optical response of NLSMs ZrSiS and ZrSiSe in the near-infrared using transient reflectivity reveal two responses with different time scales. The first decays after hundreds of femtoseconds whereas the other lasts for nanoseconds which were explained as a sudden change of the electronic properties such as an increase in the electronic screening or the reduction of the plasma frequency followed by an increase of the Drude scattering rate \cite{kirby}. 

Although some ultrafast optical studies have been reported which are based on transient reflectivity and provide some information about the relaxation dynamics, a clear understanding of how the excited electrons relax and what is the physical process by which it occurs is yet to be  understood. Electron cooling can be probed by using time-resolved pump-probe spectroscopy and the relaxation dynamics can be used to shed light on the electron-electron interaction, electron-phonon interaction etc.  To the best of our knowledge, the  tr-ARPES response of ZrSiS has not been reported yet. 
These ultrafast pump-probe spectroscopy experiments are useful for observing materials perturbed far from the equilibrium state, for resolving processes which occur very fast and to control the optical and electronic properties \cite{Weber}.

In this letter, we report the ultrafast relaxation of optical and acoustic phonons in a celebrated nodal line semimetal ZrSiS. We observe slow relaxation decay of bulk nodal line in contrast to Dirac-like surface state. Our theoretical modeling based on optical and acoustic phonon cooling matches well with the observed experimental findings which reveals that optical phonon cooling is always dominant at higher temperatures which results in a fast relaxation decay whereas a sharp crossover from the optical phonon cooling to acoustic phonon cooling takes place at lower temperature resulting in slower relaxation decay. Concomitantly, both relaxation processes adhere linear decay. This work could provide  the simplistic approach to understand the relaxation dynamics in other nodal-line semimetals. 

High-quality single crystals of ZrSiS were grown by the vapor transport method. The details of the preparation method are described elsewhere \cite{Wang, Singha, Sankar}. The relaxation dynamics of the nodal-line bulk state and the surface state of ZrSiS was studied using a novel Yb:KGW amplifier-based tr-ARPES setup with an ultrashort extreme ultraviolet (XUV) source of 21.8 eV. The high photon energy of 21.8 eV is very important for accessing a relatively large momentum area since the Dirac-like surface state is near the edge of the BZ. The pump fluence for all the measurements is kept approximately at $760 \; \mathrm{\mu J / cm^{2}}$. A schematic illustration of the geometry of our tr-ARPES experiment is shown in Fig. 1(a) and the details of the experimental setup is given in the reference \cite{Liu}. 

ZrSiS crystallizes in the tetragonal crystal structure with space group P4\textit{/nmm} and is presented in Fig. 1(b). The Fermi surface measured using a photon energy of 30 eV is shown in Fig. 1(c). A diamond-shaped Fermi pocket around the $\Gamma$ point and an elliptical pocket around the X point of the Brillouin zone are observed. In the leftmost panel of Fig. 1(d), we present the Fermi surface obtained using our tr-ARPES setup and the right panels in fig. 1(d) show the constant energy contours at various binding energies equal to 280, 650, and 900 meV, respectively. The Fermi surface obtained from tr-ARPES is consistent with the Fermi surface obtained from our synchrotron results as shown in Fig. 1 (c) thus suggesting that XUV-based tr-ARPES is good enough for the detailed characterization of the ultrafast dynamics in ZrSiS. ZrSiS possesses a bulk nodal-line crossing around the $\Gamma$ point which arises from the p$_x$ and p$_y$ orbitals in the Si square net and the nearby Zr-$d$ orbitals \cite{Schoop}. The band structure of the material also exhibits Dirac crossing at the X point which is protected by the nonsymmorphic symmetry of the crystal structure \cite{fu}. 

\begin{figure}
    \includegraphics[width=\textwidth,angle=0]{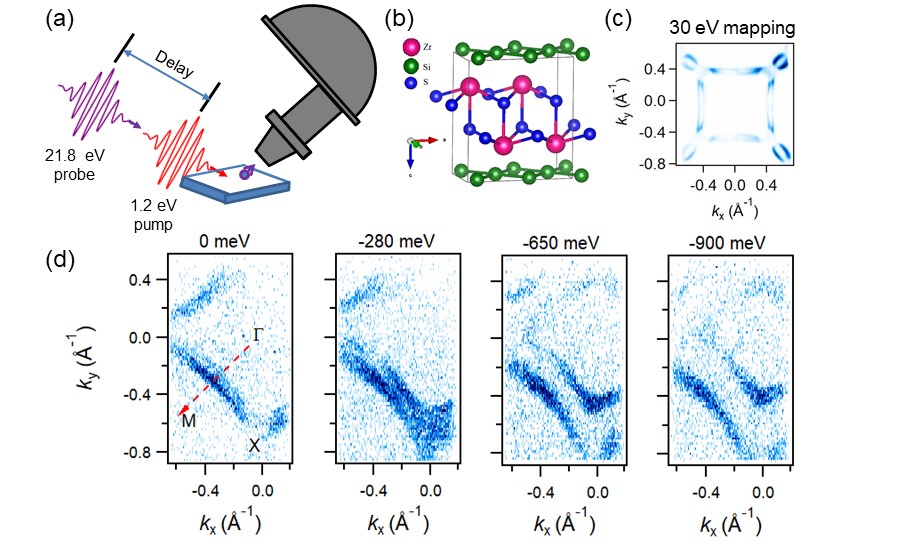}
   \caption{(a) Schematic view of pump-probe ARPES setup. The photon energies of the pump and probe pulse are 1.2 eV and 21.8 eV, respectively. (b) Tetragonal crystal structure of ZrSiS. (c) Measured Fermi surface map of ZrSiS using a photon energy of 30 eV. It consists of two diamond shaped pockets formed by the crossing of the bulk valence and conduction bands and an additional pocket at the X point formed by the surface state. (d) Fermi surface (leftmost panel) and constant energy contours at various binding energies as noted on top of the contours. The measurements in panel (d) were performed in our home-built static ARPES system.}\label{Fig_1}
\end{figure}

In the following, we will focus on discussing the ultrafast dynamics of ZrSiS.  A pump pulse with a photon energy of 1.2 eV is used to excite the sample and  the transient electronic structure is probed along the $\Gamma$-M direction by an ultrashort XUV pulse of 21.8 eV as shown in Fig. 2. Figure 2(a) and (b) show the bulk nodal-line state at two different time delays.  Figure 2(a) is measured at a delay of -6.3 ps, which means the probe pulse is 6.3 ps earlier than the pump pulse. The measured linearly dispersing nodal-line bulk bands are consistent with the previous study \cite{Neupane}, and reasonably, no photoexcited states are observed above the Fermi level. When the delay time between the pump and the probe pulse is changed to 0.4 ps, the electrons are excited to the unoccupied states above the Fermi level by the pump pulse and the transient electronic structure is probed by the ultrashort XUV pulse, which is illustrated in Fig. 2(b).
 
 \begin{figure}
\centering
    \includegraphics[width=0.9\textwidth,angle=0]{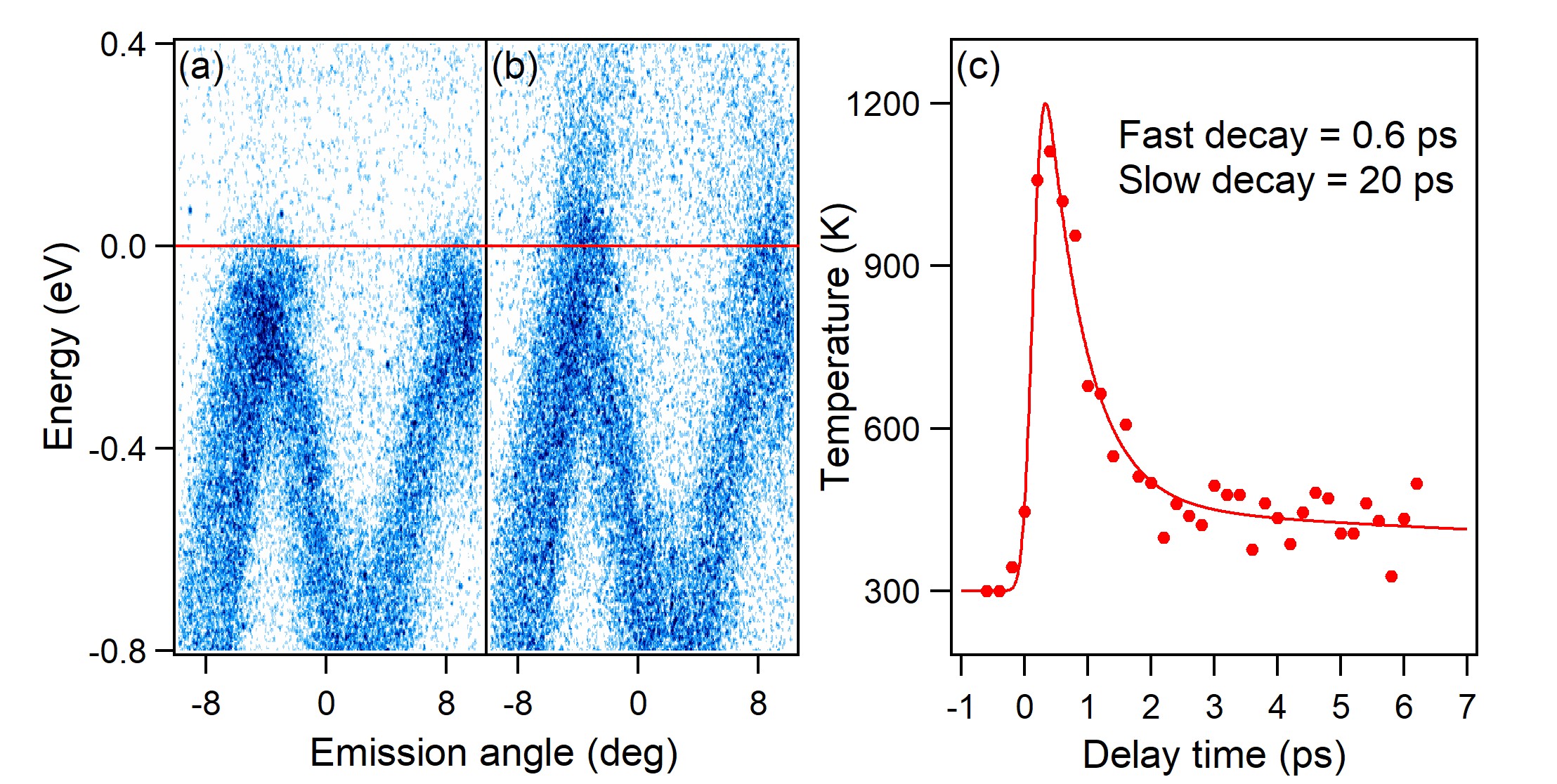}
    \caption{tr-ARPES measurements of the nodal-line bulk states. Experimental band dispersion measured along the $\Gamma$-M direction at delay times of (a) -6.3 ps  and (b) 0.4 ps. The unoccupied non-equilibrium states can be observed only when the delay value is positive. (c) The transient electronic temperature as a function of the time delay ( dotted line) and the corresponding fitting curve (solid line). The fitting function is the convolution between a Gaussian pulse and a bi-exponential decay. 
}\label{Fig_2}
\end{figure}
 
In order to systematically study the ultrafast dynamics of the hot electrons of the nodal-line bulk states, we proceed to analyze the transient electronic distributions. First, the temperature of the hot electrons over the time delay is analyzed by integrating the momentum of the ARPES spectra to get the energy distribution curves (EDCs), and then extracting the electronic temperatures by fitting the EDCs with the Fermi-Dirac distribution as a function of the time delay, which is shown as red dots in Fig. 2(c). At negative delays, the electronic temperature stays at around 300 K since the measurement is performed at room temperature. Upon excitation, the electronic temperature suddenly increases to around 1200 K at 0.4 ps, then drops to around 450 K at 6 ps. The structure of the measured transient electronic temperature can be considered as containing three parts: a very sharp increase (0 - 0.4 ps), a fast decay (0.4 - 2 ps), and another slow decay (2 - 6 ps). The transient temperature is fitted with a convolution between a Gaussian pulse and a bi-exponential decay, which is shown as the red line in Fig. 2(c). From the fitting curve, the fast and the slow decays can be extracted, which are 0.6 ps and 20 ps, respectively.

\begin{figure}
\centering
    \includegraphics[width=0.9\textwidth,angle=0]{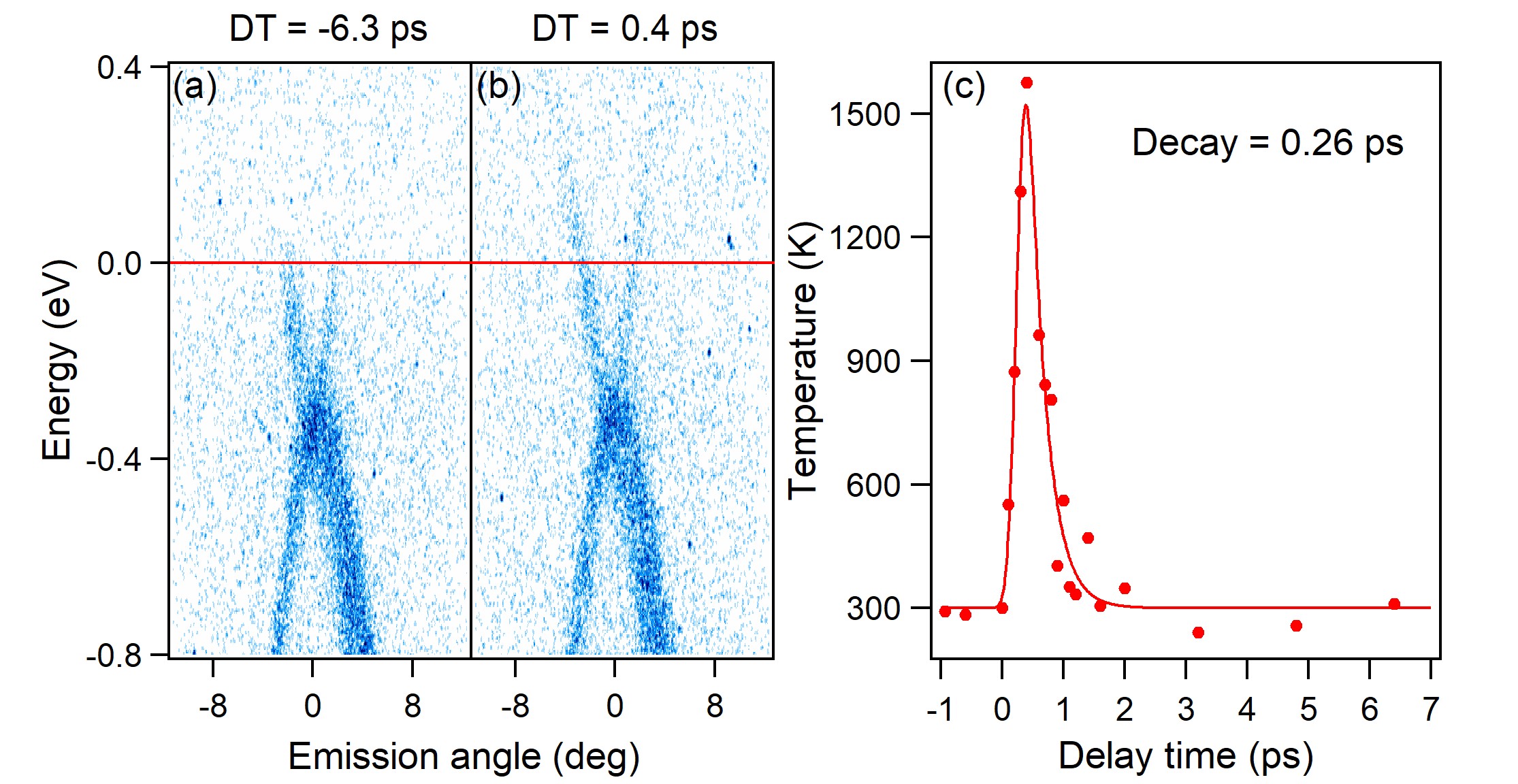}
    \caption{tr-ARPES measurements of the Dirac-like surface states.  Experimental band dispersion measured along the X-M direction at delay times of  (a) -6.3 ps  and (b) 0.4 ps. (c) The transient electronic temperature as a function of the time delay (red dots) and the corresponding fitting curve (red line). The fitting function is the convolution between a Gaussian pulse and a single exponential decay.}\label{Fig_3}
\end{figure}

\begin{figure}
\centering
    \includegraphics[width=0.9\textwidth,angle=0]{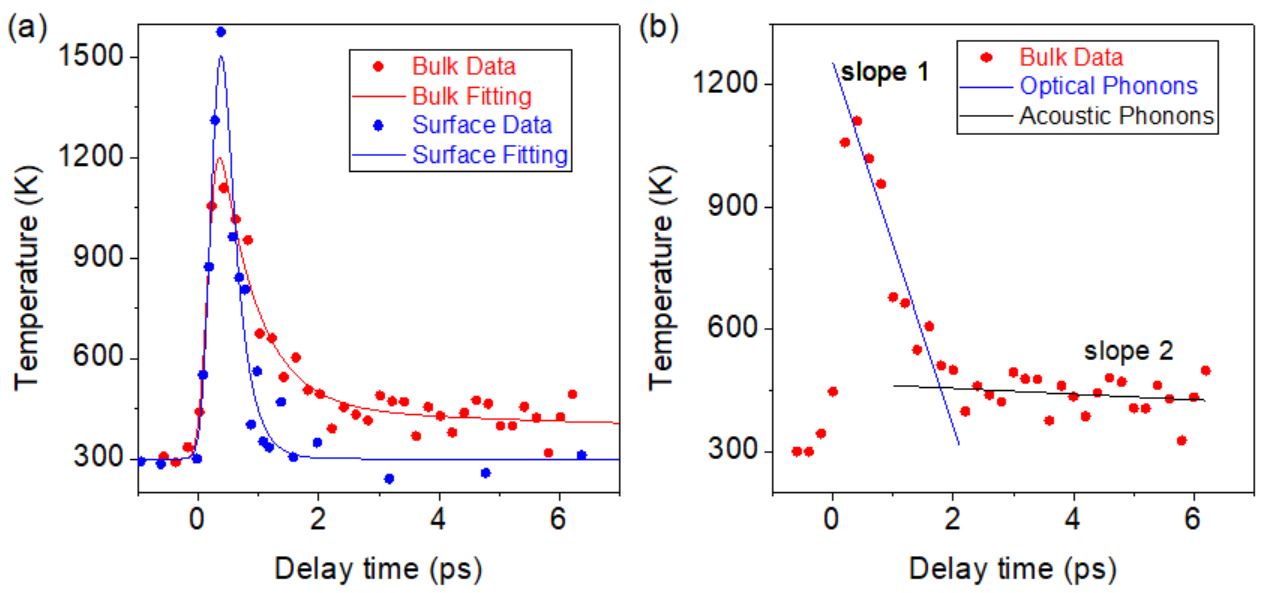}
    \caption{(a) Comparison between the transient electronic temperature of the nodal-line bulk and the surface states. The measured temperature data are shown as dots, with the convolution fitting shown as lines. The decay time of the nodal-line bulk state is found to be longer than the surface state. (b) The measured bulk temperature are shown as red dots. The blue and black lines, which are obtained from our theory, represent the optical and acoustic phonons, respectively.}\label{Fig_4}
\end{figure}

In order to understand the transient dynamics of the Dirac-like surface states, we have performed ultrafast pump-probe spectroscopy measurements along the X-M direction as shown in Fig. 3. Figures 3(a) and (b) show the ARPES spectra measured at different time delays between the pump and probe pulses. Fig. 3(a) shows the ARPES spectra before the pump pulse excites the electrons and hence no unoccupied states are observed above the Fermi level. However, in Fig. 3(b), we can see the upper region of the Dirac cone is populated at a positive time delay of 0.4 ps after photoexcitation. The Dirac point is found to be around 280 meV below the Fermi level, which is consistent with the previous result \cite{Neupane}. To analyze the response of the photo-induced perturbation in this Dirac-like state, we integrated the momentum of the ARPES spectra to obtain the experimental EDCs which was fitted by considering the spectral function as a representation of the Fermi-Dirac distribution function for various time delays to extract the electronic temperatures. The resulting electronic temperature is shown as red dots in Fig. 3(c), which was fitted by a convolution between a Gaussian pulse and a single exponential decay as shown by red line. From the fitting curve, we can extract a decay time of 0.26 ps. 

In order to compare the ultrafast relaxation dynamics between the nodal-line bulk state and the Dirac-like surface state, we plotted the transient electronic temperature as a function of the delay time in Fig. 4(a). We observe that the temperature increases rapidly for both the bulk and the surface states, soon after the photoexcitation of the electrons due to the pump pulse. Interestingly, the relaxation dynamics in the cooling process is quite different. The relaxation time for the Dirac-like surface state is around $\sim$ 0.26 ps, which is equivalent or smaller than the time resolution of our setup. We can easily attribute this to the electron-electron interaction process \cite{Gatti}. For the nodal-line bulk state, there are two prominent features. One is the fast relaxation decay with a time scale of $\sim$ 0.6 ps, the other is a persistent state, with decay constant of around $\sim$ 20 ps. 

To understand the underlying physics of the unique features of the transient nodal-line bulk state, we compute the temperature relaxation due to optical and (longitudinal) acoustic phonons using a two-band model of a nodal-line semimetal. The nodal-line is a ring of radius $Q$ located in $k_x$-$k_y$ plane in momentum space. The energy-momentum relation of electrons near the nodal-line is 

\begin{align}
\epsilon_{n\v{k}} = nv\hbar[k_z^2 + (k- Q)^2]^{1/2}
\end{align}
where in cylindrical coordinates $\v{k}=(k,\phi_k,k_z)$, $n=\pm 1$ denotes conduction $(+1)$ and valence (-1) bands respectively, $v$ is the velocity of Dirac nodal quasiparticles, $k$ is the momentum in the radial direction. The relaxation model is

\begin{align}
\frac{d\mathcal{E}}{dt} =-\frac{1}{V}\sum_{n\v{k}} \epsilon_{n\v{k}} \frac{d f_{n\v{k}}}{dt},
\end{align} 
Where $\mathcal{E}$ is the internal energy of eletrons and
\begin{align}
\frac{d f_{n\v{k}}}{dt} &=  \sum_{m\v{p}}\big[f_{n\v{k}}(1\hspace{-2pt}- \hspace{-2pt} f_{m\v{p}}) W_{n\v{k},m\v{p}} \hspace{-2pt} -\hspace{-2pt}(n\v{k} \leftrightarrow m\v{p})\big],\\
W_{n\v{k},m\v{p}} =&  \frac{2\pi}{\hbar} \sum_{\v{q}} M_{\v{q}} [(N_{\v{L}} \hspace{-2pt}+\hspace{-2pt} 1)\delta_{\v{k},\v{p}+\v{q}} \delta(\epsilon_{n\v{k}}\hspace{-2pt}-\hspace{-2pt}\epsilon_{m\v{p}}\hspace{-2pt}-\hspace{-2pt}\hbar\omega_{\v{q}}) \nn \\
&~~~~~~~~~~~+ N_{\v{L}} \delta_{\v{k},\v{p}-\v{q}}\delta(\epsilon_{n\v{k}}\hspace{-2pt}-\hspace{-2pt}\epsilon_{m\v{p}}\hspace{-2pt}+\hspace{-2pt}\hbar\omega_{\v{q}}]
\label{eq:Wmatrix_el}
\end{align}
$M_{\v{q}}= \hbar^2 D^2 q^2 (1 + s_{nm}\cos\theta)/4\rho V \hbar \omega_{\v{q}}$ is the amplitude of phonon-electron scattering, $\v{q}$ is the phonon momentum, $\omega_{\v{q}}$ is the phonon dispersion relation, $D$ is the deformation potential of acoustic phonons, $\rho$ the ion mass density, $V$ is the volume, $\theta$ the angle between $\v{k}$ and $\v{p}$, and $s_{nm}=1$ for intraband and $-1$ for interband scattering. To obtain an analytical expression we evaluate the collision integral to lowest order in the small $c/v$  limit where $c$ is the sound speed. We also assume the nodal radius is the largest momentum scale and that the lattice temperature $T_L$ is larger than Bloch-Gruneisen temperature. 

In the regime applicable to our experiment, we find a linear temperature relaxation due to optical phonons. We have fitted the experimental data in Fig. 4(b) with the equation  
\begin{align}
k_B T_e =k_B T_0 - \hbar\omega_0 \gamma_{o} t,
\label{eq:relax_nlsm_over_case2}
\end{align}
obtaining the slope 1. Here $\omega_0$, $T_e$, $T_0$  and $T_L$ are the optical phonon energy, electron temperature, initial electron temperature (1200 K), lattice temperature, respectively and  $\gamma_{o}=2\times3 g^2 Q \omega_0^2 (1/6 + 2\bar{\mu}^3/3)/ \pi^2 v^2 \hbar \mu$ where  g = electron-optical phonon coupling, $\mu$ is the chemical potential (see SM \cite{SM} for more details). The slope 2 in Fig. 4(b) is fitted due to relaxation associated with the acoustic phonons using the equation 
\begin{align}
k_B T_e = k_B T_0 - \gamma_{a} t
\label{eq:doped_rel}
\end{align}
where $\gamma_{a} = 2\times 3 D^2 Q^3\epsilon_F /4 \pi^2 \rho \hbar v^2$  (See SM for more details \cite{SM}). Using the parameters as shown in Table S1 (see SM \cite{SM}), the slope 1 and slope 2 as shown by solid lines in Fig. 4(b) obtained from theoretical fits are -439.6 K/ps and -7.07 K/ps, respectively. This is in excellent agreement with the experimental fit for both slope 1 and slope 2 as -446.44 K/ps and -7 K/ps, respectively. This suggests that when the electronic temperature is high, the acoustic phonon cooling process is strongly suppressed, and optical cooling plays a major role, which results in a fast linear relaxation decay. When the Fermi energy is close to the nodal-line, relaxation due to acoustic phonon is strongly suppressed for all the temperatures above the lattice temperatures. This is because the nodal ring constrains the acoustic phonons to a very narrow region of momentum space. However, when the electronic temperature is small (below 500 K), the acoustic cooling process becomes dominating, which leads to a much slower relaxation decay. 

In conclusion, we have performed XUV-based tr-ARPES measurements on a topological nodal-line semimetal ZrSiS. The detailed analysis of the transient ARPES spectra as a function of the time delay suggests the ultrafast relaxation of the hot electrons in the nodal-line bulk state is much longer than the surface states. In addition to this, for the  nodal-line bulk states, we find that at high electron temperatures the optical-phonon relaxation time scale is much shorter than the time scale corresponding to acoustic phonons and may be evidence of a more general mechanism in TNSMs. Our results provide first insights into electron cooling process for the nodal-line bulk state via energy exchange with the lattice.

\indent ACKNOWLEDGMENTS.

M.N. acknowledges support from the Air Force Office of Scientific Research under Award No. FA9550-17-1-0415 and the Air Force Office of Scientific Research MURI (FA9550-20-1-0322). B.M.F. acknowledges support from NSF grant DMR-2015639 and DOE-NERSC under contract DE-AC02-05CH11231. M.C. acknowledges support from the Air Force Office of Scientific Research under Award Numbers FA9550-16-1-0149 and FA9550-20-1-0284. J.E.B.'s current address is Department of Chemistry, University of California Berkeley.\\

$^\dagger$ Corresponding author: Madhab.Neupane@ucf.edu, benjamin.fregoso@gmail.com

\end{document}